\documentclass{article}
\usepackage[utf8]{inputenc}
\usepackage[lmargin=0.65in,rmargin=0.65in,tmargin=0.9in,bmargin=0.8in]{geometry}
\usepackage{hyperref,lmodern,textcomp,graphicx}

\usepackage[table]{xcolor} \usepackage{caption,fancyhdr}
\usepackage{color,amsmath,amssymb,mathpazo,mathtools}
\usepackage{vwcol}
\usepackage{lipsum}
\usepackage{tgpagella}
\usepackage{indentfirst}
\usepackage{bchart}
\usepackage{lettrine,multicol}
\usepackage[sorting=none,style=chem-angew,autocite= superscript]{biblatex}
\newcommand{\HA}[1]{{\color{black} #1}}

\usepackage{tcolorbox}%
\definecolor{mycolor}{rgb}{0.122, 0.435, 0.698}
\makeatletter
\newcommand{\mybox}[1]{%
  \setbox0=\hbox{#1}%
  \setlength{\@tempdima}{\dimexpr\wd0+13pt}%
  \begin{tcolorbox}[boxrule=0pt,arc=0pt,
      left=6pt,right=6pt,top=6pt,bottom=6pt,boxsep=1pt,width=\textwidth]
    #1
  \end{tcolorbox}
}

\fancypagestyle{firstpage}
{\fancyhead[L]{\textsc {\color {gray} {\large }}}    
\fancyhead[R]{\large \textsc{\color {gray} {Research Article}}}}
    
\graphicspath{{Figures/}}

\begin{document}

\thispagestyle{firstpage}
{\noindent \LARGE \textit{Active elastic metamaterials with equidistant solely resonant bandgaps}} \\ [0.5em] 

\noindent {\large \textit{Hasan B. Al Ba'ba'a}$^{\ddagger}$} \\

\noindent \begin{tabular}{c >{\arraybackslash}m{6in}}
    \includegraphics[]{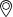} &
    \noindent {\small Department of Mechanical Engineering, Union College, Schenectady, NY 12308, USA} \\[0.25em]
    
    \includegraphics[]{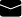}& \noindent {\href{mailto:albabaah@union.edu}{\small albabaah@union.edu}$^{\ddagger}$}\\
\end{tabular}


\mybox{\textit{Elastic metamaterials are man-made structures with properties that transcend naturally occurring materials.~One predominant feature of elastic metamaterials is locally resonant bandgaps, i.e., frequency ranges at which wave propagation is blocked.~Locally resonant bandgaps appear at relatively low frequency and arise from the existence of periodically placed mechanical local resonators.~Typically, elastic metamaterials exhibit both locally resonant and Bragg-scattering bandgaps, which can generally be different in width and frequency ranges.~This paper proposes two designs of active elastic metamaterials that only exhibit locally resonant bandgaps, which are infinite in number, evenly spaced in the frequency spectrum, and identical in width.~The mathematical model is established using the transfer matrix method and synthesis of locally resonant bandgaps is achieved via an active elastic support with carefully designed frequency-dependent stiffness.~A single unit cell of each proposed metamaterials is thoroughly studied, and its dispersion relation is derived analytically, along with the periodically repeating bandgap limits and widths.~Following the dispersion analysis and bandgap parametric studies, finite arrays of the proposed metamaterials are considered, and their frequency response is calculated to verify the analytical predictions from dispersion analyses.}}

\vspace{0.25cm}
\noindent \textbf{Keywords
}

\noindent Locally resonant bandgaps; active elastic metamaterials; equidistant bandgaps; frequency-dependent stiffness.

\noindent\rule{7.2in}{0.5pt}
\section{Introduction}
\HA{Since their inception in the early 2000s \cite{liu2000locally}, locally resonant materials, often referred to as elastic metamaterials (EMMs), have witnessed a surge in research interest, thanks to the emergence of locally resonant bandgaps~\cite{Hussein2014,liu2020review,wang2020tunable,lee2023elastic,CONTRERAS2024117663}}.~By definition, bandgaps are frequency ranges where wave propagation is effectively blocked, without the need for any traditional dissipative mechanisms.~Mitigation of waves within locally resonant bandgaps is purely a consequence of local-resonance effects \cite{albabaa2017PZ} (or negative effective mass behavior \cite{huang2009negative}), unlike the Bragg-scattering phenomenon that is caused by destructive interference of the incident and reflected waves~\cite{Hussein2014}.~Locally resonant bandgaps are superior to those \HA{arising from} Bragg-scattering, owing to their independence on the unit-cell size and considerably low operational frequencies at the sub-wavelength regime \cite{huang2011study,Zhu2014,Baravelli2013}.~\HA{Incorporating inerters (i.e., ideal mechanical elements with mass amplification capabilities~\cite{smith2002synthesis}) in EMMs have also demonstrated an ability of exhibiting locally resonant bandgaps at even lower frequency ranges \cite{jamil2022inerter}.~Consequently, EMMs have been at the forefront of wave propagation research, enabling a multitude of potential applications that include amplification of damping due to metadamping effects \cite{hussein_meta,depauw2018metadamping,bacquet2018metadamping,bacquet2018dissipation,al2021metadamping}, enhanced energy harvesting \cite{lin2021piezoelectric}, subsurface flow-stabilization \cite{kianfar2023phononic}, and negative refraction and elastic wave lens~\cite{quadrelli2024subwavelength}, among others \cite{lee2023elastic,CONTRERAS2024117663}}.

A common configuration of EMMs is built via a host elastic structure with periodically placed (and identical) mechanical local resonators of relatively low stand-alone natural frequencies to enable sub-wavelength locally resonant bandgaps.~Furthermore, EMMs may be equipped with different resonator properties along the host elastic structure, advantageous for attenuating waves with multi-frequency harmonics~\cite{Pai2010} or widening bandgap range \HA{\cite{celli2019bandgap,stein2022widening,meng2023theoretical}}.~While EMMs are mainly designed to exhibit single or multiple locally resonant bandgaps, Bragg-scattering bandgaps can still arise, and the conditions leading to the transition from one to the other have been established \cite{liu2012wave}.~As a result, various studies have focused on combining Bragg-scattering and locally resonant bandgaps and derived the conditions leading to such phenomenon and their relationship to structural natural frequencies as well as Bragg-conditions~\HA{\cite{gao2022ultrawide,cleante2022formation,krushynska2017coupling,hussein_AIP,xiao2012longitudinal,xiao2011formation}}.~Another recent endeavor in this domain has been focused on identifying patterns in bandgap formation (periodic versus non-periodic bandgap widths/location) and conditions leading to bandgap closure for phononic crystals undergoing axial vibrations~\cite{Albabaa2023Patterns}.~Yet, a framework that enables infinite number of periodic locally resonant-only bandgaps (i.e., no Bragg-scattering bandgaps) that are equally spaced and having identical patterns of bandgap and propagation-zone widths remains lacking. 

In this paper, two designs of active EEMs are introduced, with the capability of exhibiting an infinite number of locally resonant bandgaps that are identical in width, equidistant, and periodically repeating. The unique characteristics of the proposed designs are achieved via an elastic continuum rod with periodically placed elastic supports of frequency-dependent stiffness, designed specifically to realize the desired functionality. To engineer the bandgap patterns and dispersion behaviors, the dispersion relations are derived analytically using the transfer matrix method. A single unit cell of the active EMMs is thoroughly studied, and the periodic bandgap limits and widths are derived analytically.~Finally, the analytical dispersion predictions are verified using the frequency response of the proposed active EMMs with a finite number of unit cells.

\section{Design of Active Elastic Metamaterials}
\subsection{Background: discrete elastic metamaterials and diatomic lattices}

\begin{figure*}[]
     \centering
\includegraphics[]{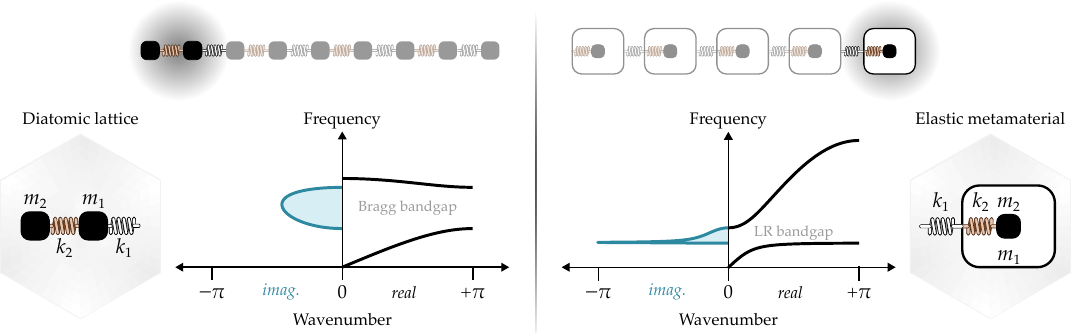}
\caption{Schematics of a typical discrete diatomic lattice (left) and EMM (right), showing the inertial and stiffness properties for the self-repeating unit cell.~Typical dispersion relations of both cases are shown in the figure for reference.~Bragg-scattering and locally resonant~(LR) bandgaps are observed in the diatomic lattice and EMM, respectively, and defined by the shaded regions with non-zero imaginary wavenumbers.}
\label{fig:sch2}
\end{figure*}

\begin{table*}[]
\caption{Expressions for the variable $\alpha$ in Equation~(\ref{eq:discrete_disp}) and functions $\Phi_n$ and $\Phi_d$ of the driven-wave formulation in Equation~(\ref{eq:q_driven}) for both the diatomic lattice and EMM models. For an interpretation of the mass and stiffness parameters of both systems, see Figure~\ref{fig:sch2}.}
\centering
\begin{tabular}{l l l}
\hline
Parameter & Diatomic Lattice & Elastic Metamaterial \\
\hline
$\alpha$ & $(m_1 + m_2)(k_1+k_2)$ & $(m_1+m_2)k_2 + 2m_2 k_1 (1-\cos(q))$ \\
\hline
$\Phi_n$ & $m_1 m_2 \omega^4 - (m_1+m_2)(k_1 + k_2) \omega^2 + 2 k_1 k_2$ & $m_1 m_2 \omega^4 - \left( (m_1+m_2)k_2 + 2m_2k_1  \right) \omega^2 + 2 k_1 k_2$ \\
\hline
$\Phi_d$ & $2 k_1 k_2$ & $2 k_1 (k_2-m_2 \omega^2)$ \\
\hline
\end{tabular}
\label{table:functions}
\end{table*}

To demonstrate the design methodology of the proposed active EMMs, it is essential to first revisit the dispersion relations of the discrete model of EMM and a diatomic lattice. Figure~\ref{fig:sch2} details the inertial and stiffness properties for both models, which generally have the following dispersion relation \cite{albabaa2017PZ,albabaa2017PC}:
\begin{equation}
    m_1 m_2 \omega^4 - \alpha \omega^2 + 2k_1k_2(1-\cos(q)) = 0
    \label{eq:discrete_disp}
\end{equation}
where $q$ is the non-dimensional wavenumber, $\omega$ is the excitation frequency, and the parameter $\alpha$ is defined in Table~\ref{table:functions} for both models.~Rearranging the dispersion relations into the driven-wave formulation, i.e., the input being the frequency~$\omega$ and the output being wavenumber $q$ (having a complex quantity within bandgaps), one can show that:
\begin{equation}
    q = \cos^{-1}\left(\frac{\Phi_n}{\Phi_d} \right)
    \label{eq:q_driven}
\end{equation}
The functions $\Phi_n$ and $\Phi_d$ for the diatomic lattice and the discrete EMM are listed in Table~\ref{table:functions}.~Interestingly, $\Phi_n$ becomes identical for both models if the constitutive masses are identical, i.e., $m_1 = m_2$. On the other hand, the denominator function $\Phi_d$ differs between the two cases as it is constant and frequency-dependent for the diatomic lattice and discrete EMM, respectively. As a consequence, and unlike the diatomic lattice case, EMM exhibits a locally resonant bandgap around the zero of $\Phi_d$, which occurs precisely at the stand-alone natural frequency of the local resonator. In extension, having identical masses ($m_1 = m_2$) and stiffnesses ($k_1 = k_2$) also yields identical $\Phi_n$ functions for both cases. In such a scenario, the diatomic-lattice's dispersion becomes that of the monatomic lattice with no bandgap.~However, a locally resonant bandgap still opens in EMM even if the inertial and stiffness properties are equal, given that a zero of the function $\Phi_d$ still exists at the local resonator's natural frequency.~In conclusion, it is possible to artificially introduce a function $\Phi_d$ with zeros at the desired frequency(ies) to induce locally resonant bandgap(s) around them when combined with a function $\Phi_n$ of a monatomic lattice or a uniform elastic rod as will be elaborated next. 

\begin{figure*}[t]
     \centering
\includegraphics[]{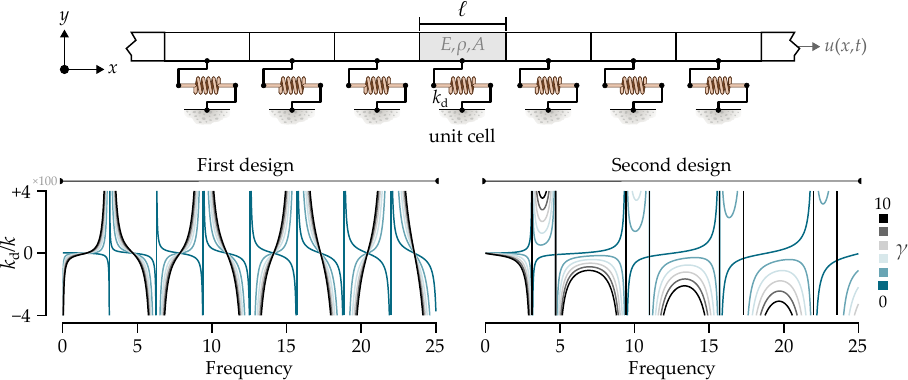}
\caption{(\textit{top})~Schematic of an elastic rod with a unit-cell length of $\ell$ and equally spaced elastic foundation $k_\text{d}$.~The continuous elastic rod (of Young's modulus $E$, density $\rho$, and cross-sectional area $A$) undergoes longitudinal vibrations, quantified by $u(x,t)$ as functions of axis $x$ and time $t$.~(\textit{bottom}) Frequency response of the frequency-dependent stiffness $k_\text{d}$ relative to the effective stiffness of the continuum rod segment, $k = EA/\ell$, at different values of the tuning parameter~$\gamma$.}
\label{fig:sch}
\end{figure*}

\subsection{First design: identical, equally spaced and symmetric bandgaps}

Consider a continuum elastic rod, with a unit-cell length of $\ell$, a cross-sectional area of $A$, and a sonic speed of $c = \sqrt{E/\rho}$, where $E$ and $\rho$ are Young's modulus and density of the rod, respectively.~Considering longitudinal waves, the dispersion relation is linear, and it can be written in the following non-dimensional form:
\begin{equation}
    \Omega = q
\label{eq:linear_dispersion}
\end{equation}
where $\Omega = \omega \ell/c$ is a normalized version of the excitation frequency $\omega$. Equation~(\ref{eq:linear_dispersion}) can be re-cast into an analogous form to that in Equation~(\ref{eq:q_driven}) and it immediately follows that $\Phi_d = 1$ and $ \Phi_n = \cos(\Omega)$.~To enable periodic, identical, and equally spaced locally resonant bandgaps, I propose the formula $\Phi_d = \sin(\Omega)/\gamma$, which, based on Equation~(\ref{eq:q_driven}), results in the following driven-wave dispersion relation:
\begin{equation}
    q = \cos^{-1} \left(\frac{\gamma \cos(\Omega)}{\sin(\Omega)} \right)
    \label{eq:best_disp_rel}
\end{equation}
where $\gamma $ is a tuning parameter for the bandgaps' limits. Having the term $\sin(\Omega)$ in the denominator $\Phi_d$ means that evenly spaced (and infinite in number) local resonances emerge at $\Omega = n\pi$, where $n \in \mathrm{N}_0$ and $\mathrm{N}_0$ denotes all natural numbers including zero. Equivalently, the free-wave dispersion branches can be derived by rearranging the terms in Equation~(\ref{eq:best_disp_rel}), resulting in
\begin{equation}
     \Omega= n\pi + \tan^{-1} \left(\frac{\gamma}{\cos(q)} \right)
    \label{eq:2nd_best_disp_rel}
\end{equation}
From the solutions of Equation~(\ref{eq:2nd_best_disp_rel}) at $q = 0, \pi$, lower and upper bandgap limits, denoted here as $\Omega_l$ and $\Omega_u$, respectively, can be found using the following equations for all $n\geq 1$:
\begin{subequations}
    \begin{equation}
    \Omega_l = n \pi - \tan^{-1}(\gamma)
\end{equation}
\begin{equation}
    \Omega_u = n \pi + \tan^{-1}(\gamma)
    \label{eq:bglimits_d1_u}
\end{equation}
\label{eq:bglimits_d1}
\end{subequations}
As inferred from the structure of the bandgap limits formulae in Equation~(\ref{eq:bglimits_d1}), they are of equal distance of $\tan^{-1}(\gamma)$ from the local resonance frequencies at $n \pi$.~As a result, the bandgap limits are perfectly periodic and infinite in number, with identical widths of:
\begin{equation}
    \Delta \Omega = 2 \tan^{-1}(\gamma)
\end{equation} Note that the lower limit of the very first locally resonant bandgap is equal to $\Omega = 0$ (referred to here as \textit{zero-frequency bandgap}) and its upper limit being $\Omega_0 = \tan^{-1}(\gamma)$, derived from substituting $n=0$ in Equation~(\ref{eq:bglimits_d1_u}).~Such a zero-frequency bandgap has half the width of the rest of the bandgaps, and its upper limit also signifies a lower cutoff frequency of this first design of active EMM.

To achieve the dispersion relation in Equation (\ref{eq:best_disp_rel}), a grounded elastic element of stiffness $k_\text{d}$ (\HA{similar to the dynamics of EMMs with branched mechanical resonators~\cite{bastawrous2021theoretical}}) is attached to the rod at the middle of the unit cell, as shown in Figure~\ref{fig:sch}. The elastic element $k_\text{d}$ is engineered to exhibit a desired frequency-dependent stiffness, enabling repeating local resonances induced by the zeros of the function $\Phi_d$.~Here, the transfer matrix method is utilized to find the function of the frequency-dependent stiffness $k_\text{d}$. The purpose of a transfer matrix $\mathbf{T}$ is to relate the state vector (displacement and internal force) from one end of a unit cell to its peer in the neighboring cell:
\begin{equation}
 \mathbf{z}_{i+1} = \mathbf{T} \mathbf{z}_{i}
\end{equation}
where $\mathbf{z}_{i}$ is the unit-cell's state vector at one end of the $i$th unit cell. To establish the transfer matrix for the unit cell in Figure~\ref{fig:sch}, the transfer matrices of the uniform elastic rod segments and ground springs are needed. It can be shown that the transfer matrix of a uniform rod for half of the unit cell length is found using the following transfer matrix: 
\begin{equation}
    \mathbf{Y} = \begin{bmatrix}
\cos(\Omega/2) & \frac{1}{k \Omega} \sin (\Omega/2) \\ -k \Omega \sin (\Omega/2) & \cos (\Omega/2)
\end{bmatrix}
\label{eq:TM_1}
\end{equation}
which is presented here in terms of the non-dimensional frequency $\Omega$, with $k = EA/\ell$ being the effective stiffness of the rod segment. For a grounded spring, on the other hand, the following transfer matrix is applied to account for its effect, which, in this case, is identical in form to that of a transfer matrix of a typical local resonator \cite{hussein_AIP}:
    \begin{equation}
\mathbf{P}=
\begin{bmatrix}
1 & 0 \\ k_{\text{d}}(\Omega) & 1
\end{bmatrix}
\label{eq:TM_2}
\end{equation}
Thus, the transfer matrix of the entire unit cell is derived by multiplying the individual transfer matrices defined in Equations~(\ref{eq:TM_1})~and~(\ref{eq:TM_2}):
\begin{equation}
    \mathbf{T} =\mathbf{Y} \mathbf{P} \mathbf{Y} = 
     \begin{bmatrix}
\cos(\Omega) + \frac{k_{\text{d}}}{2k\Omega} \sin(\Omega) & \frac{1}{2\Omega^2 k^2} \left( k_{\text{d}}\left(1-\cos(\Omega) \right) + 2\Omega k \sin(\Omega) \right)
\\ 

\frac{k_{\text{d}}}{2} \left(1+\cos(\Omega)\right) - k \Omega \sin(\Omega)
& \cos(\Omega) + \frac{k_{\text{d}}}{2k\Omega} \sin(\Omega)
\end{bmatrix}
\end{equation}
Following the development in Ref.~\cite{Albabaa2023Patterns}, the non-dimensional wavenumber $q$ is related to the transfer matrix $\mathbf{T}$ via $\cos(q) = \text{tr}(\mathbf{T})/2$, with $\text{tr}(\mathbf{T})$ being the trace of $\mathbf{T}$ and is given by:
\begin{equation}
    \text{tr}(\mathbf{T}) = 2 \cos(\Omega) + \frac{k_{\text{d}}}{k \Omega} \sin(\Omega)
     \label{eq:traceT_d1}
\end{equation}
As such, the function $k_{\text{d}}(\Omega)$ is derived via combining $\cos(q) = \text{tr}(\mathbf{T})/2$ and Equations~(\ref{eq:best_disp_rel})~and~(\ref{eq:traceT_d1}), yielding:
\begin{equation}
    k_{\text{d}}(\Omega) = -  \frac{2k\Omega}{\tan(\Omega)} \left(1 - \frac{\gamma}{\sin(\Omega)} \right)
    \label{eq:kd1}
\end{equation}

\subsection{Second design: no zero-frequency bandgap}

\begin{figure*}[h]
     \centering
\includegraphics[]{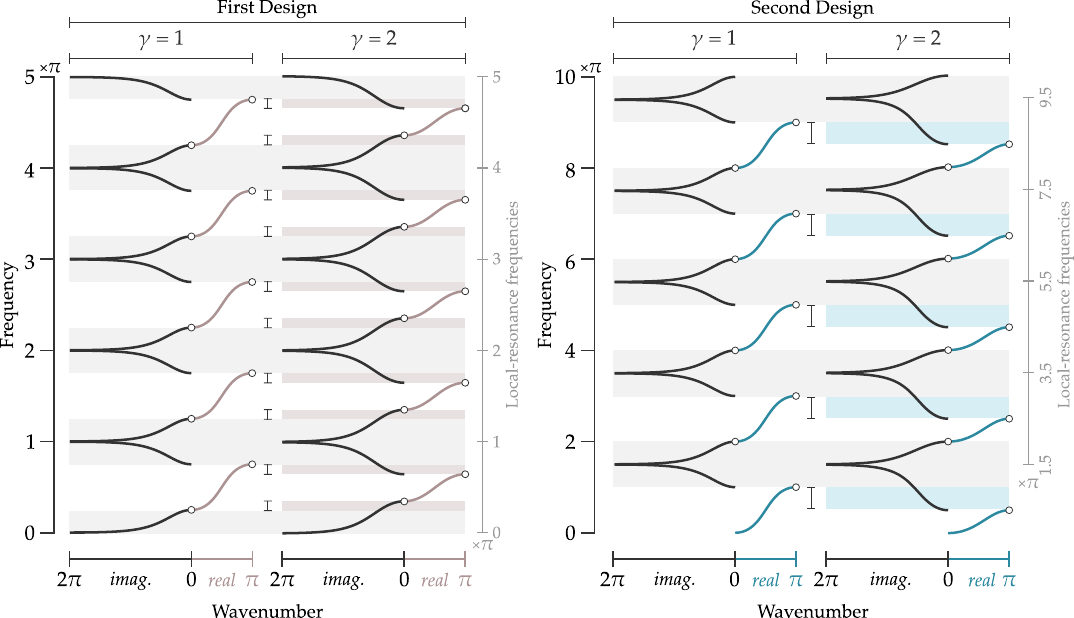}
\caption{Band structure of the two designs of active EMM with two values of the tuning parameter, namely $\gamma = 1$ and $\gamma = 2$. All cases exhibit identical locally resonant bandgaps and dispersion branches, and the bandgap width gets larger with an increase in~$\gamma$.~For the first design, and as $\gamma$ increases, the bandgap expands symmetrically about the center of the bandgap (i.e., the marked local-resonance frequencies on the right vertical axis) for all bandgaps, regardless of the value of the parameter $\gamma$, except for the zero-frequency bandgap that always starts at $\Omega = 0$. On the other hand, no zero-frequency bandgap emerges in the second design given that the first locally resonant bandgap is centered around the local-resonance frequency $\Omega = 3\pi/2$, and the behavior recurs every $2\pi$ in the frequency spectrum. As in the first case, the parameter $\gamma$ controls the bandgap width, but it is generally not symmetric about the local-resonance frequencies.~Note that the initial group velocity (slope of dispersion branch) at $\Omega = 0$ in the second design is zero, which is not typical for longitudinal waves with no zero-frequency bandgap. As a final note, the occurrence of local-resonance frequencies is evident from the drastic increase of the imaginary component of the wavenumber around them, and their location on the frequency spectrum is independent of the value of $\gamma$.}
\label{fig:disp}
\end{figure*}

For the dynamic stiffness $k_\text{d}$ function in the first active EMM design in Equation~(\ref{eq:kd1}), there exists a bandgap that opens starting at zero frequency as mentioned earlier. If such a bandgap is not desired, a small modification to the dispersion relation in Equation~(\ref{eq:best_disp_rel}) eliminates such gap, and the following engineered dispersion relation is proposed:
\begin{equation}
    q = \cos^{-1} \left(1-\gamma + \frac{\gamma \cos(\Omega)}{1+\sin(\Omega)} \right)
    \label{eq:non-zero-disp-rel-3rd-best}
\end{equation}
The function in the argument of cosine inverse in Equation~(\ref{eq:non-zero-disp-rel-3rd-best}) is chosen such that a zero wavenumber ($q = 0$) is returned when $\Omega = 0$, which mandates that the zero-frequency bandgap is closed.~Note that the local resonances, in this case, are found from the roots of $\sin(\Omega)+1 = 0$ and occur at:
\begin{equation}
    \Omega = (4n-1)\frac{\pi}{2}
    \label{eq:om_R_d2}
\end{equation}
for integer values of $n\geq1$. This second design of EMM results in solely resonant bandgaps having lower and upper limits of (for $n \in \mathrm{N}_0$):
\begin{subequations}
    \begin{equation}
    \Omega_l =  
2 \left(\pi n + \tan^{-1}\left(\frac{1}{\gamma-1} \right)\right)
\end{equation}
\begin{equation}
    \Omega_u = 2(n+1) \pi
\end{equation}
\end{subequations}
with an identical bandgap width of:
\begin{equation}
    \Delta \Omega = 2 \pi - 2 \tan^{-1}\left(\frac{1}{\gamma-1} \right)
\end{equation}
Following an identical procedure to that presented earlier for Equation~(\ref{eq:kd1}), the function for $k_\text{d}(\Omega)$ in this second design can be derived and it is given by:
\begin{equation}
    k_\text{d}(\Omega) = \frac{2k\Omega}{\sin(\Omega)} \left(1-\gamma-\cos(\Omega) \left(1-\frac{\gamma} {1+\sin(\Omega)}\right) \right)
    \label{eq:kd2}
\end{equation}
For reference, the frequency response of the dynamic stiffnesses $k_\text{d}$ for both designs, i.e., Equations~(\ref{eq:kd1}) and (\ref{eq:kd2}), are depicted in Figure~\ref{fig:sch} at different values of $\gamma$, demonstrating its frequency dependence and tunability with $\gamma$.

\begin{figure*}[]
     \centering
\includegraphics[]{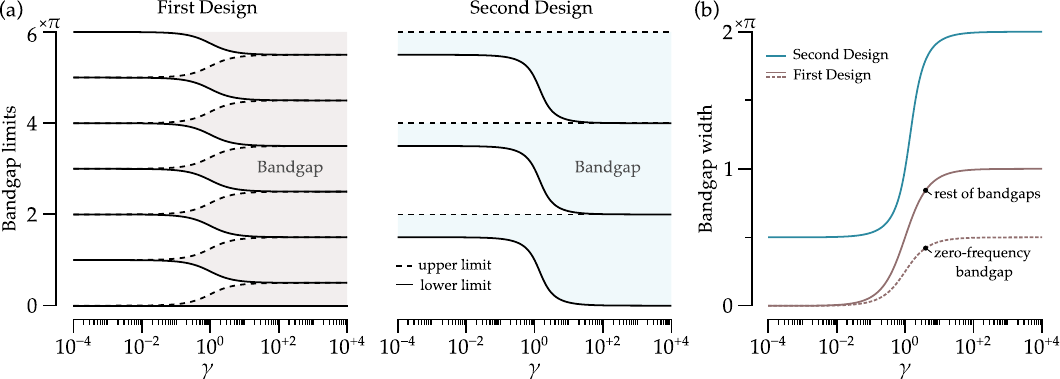}
\caption{(a)~Bandgap limits and (b)~corresponding bandgap width as a function of $\gamma$ for first and second designs of active EMM.~Shaded regions in sub-figure (a) represent the widths of the periodic bandgaps, which are quantitatively depicted in sub-figure (b).}
\label{fig:BGlimits}
\end{figure*}

\subsection{Results and discussion}
The band structures for both active EMM designs are depicted in Figure~\ref{fig:disp} for two values of $\gamma$, namely $\gamma =1$ and $\gamma =2$. As predicted earlier for the first EMM design, the first locally resonant bandgap starts at zero frequency and all bandgaps can be manipulated by changing the parameter $\gamma$. It is interesting to observe that all bandgaps grow symmetrically about the center of the bandgap as $\gamma$ increases, and the opposite is true for a decreasing value of $\gamma$ (except for the zero-frequency bandgap which grows from its upper limit). The same can be said for the second design (i.e., the dispersion in Equation~(\ref{eq:non-zero-disp-rel-3rd-best})), yet its bandgap limits are not generally symmetric about the local-resonance frequency and bandgaps grow/shrink only by changing the lower limit.~Peculiarly, the dispersion diagram of the second design exhibits a zero group velocity at the long-wave limit (i.e., at $\Omega=0$), which is not typical for longitudinal wave dispersion relations with no zero-frequency bandgap. A final difference between the band structures of the two designs is that the second one exhibits local resonances separated by $2\pi$, starting at $\Omega = 3\pi/2$, unlike the first design, which has resonances separated by a frequency range of $\pi$, starting from $\Omega = 0$. This is seen from the imaginary component of the wavenumber, where the local resonances correspond to an imaginary component of the wavenumber approaching infinity. 

The observations made regarding bandgap limits can be further confirmed by plotting them and their corresponding bandgap width as a function of $\gamma$, as seen in Figure~\ref{fig:BGlimits}. For the first design, it is observed that the bandgap limits grow apart with increasing $\gamma$, occurring symmetrically around the resonant frequencies $\Omega = n \pi$ for $n \geq 1$, as seen in the left panel of Figure~\ref{fig:BGlimits}(a). The exception here is the zero-frequency bandgap, which has a lower limit at $\Omega = 0$, confirming our observations from Figure~\ref{fig:disp}. Notice that these bandgap limits converge to odd multiples of $\pi/2$ (to multiples of $\pi$) as $\gamma \rightarrow \infty$ ($\gamma \rightarrow 0$). As in the first design, the bandgap widths in the second design increase with increasing $\gamma$, yet by pushing the lower limit down as $\gamma$ increases with the upper limit remaining constant at multiples of $2 \pi$ (Right panel of Figure~\ref{fig:BGlimits}(a)).~Examining the bandgaps' width as a function of $\gamma$, as seen in Figure~\ref{fig:BGlimits}(b), reveals that the first (second) EMM design has a maximum bandgap width of $\pi$ ($2\pi$) when $\gamma \rightarrow \infty$ and it diminishes (goes to $\pi/2$) if $\gamma \rightarrow 0$. This observation is true except for the zero-frequency bandgap in the first design, which varies between the limiting cases of 0 to $\pi/2$ corresponding to $\gamma \rightarrow 0$ and $\gamma \rightarrow \infty$, respectively. Finally, the bandgaps in the second design remain open regardless of $\gamma$, which is not the case for the first design as the bandgaps vanish at the limit of $\gamma \rightarrow 0$.

\begin{figure*}[]
     \centering
\includegraphics[]{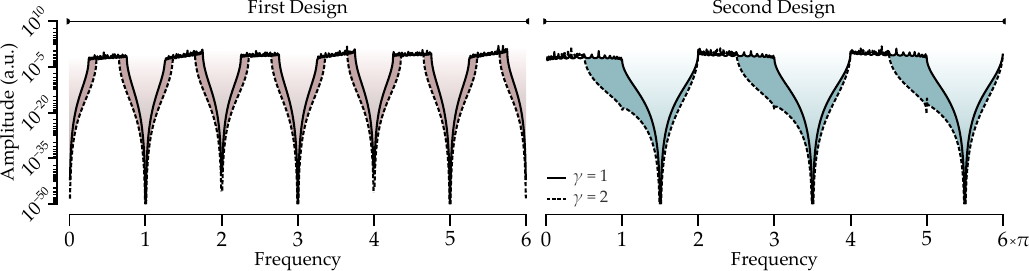}
\caption{Frequency response functions for both designs of active EMMs with 20 unit cells and $\gamma =1,2$. The rod is base-excited at one end and the displacement amplitude is measured near the opposite fixed end and normalized with respect to the base-excitation amplitude.~For the first design, the bandgaps (highlighted by the shaded regions) are symmetric around the resonant frequencies $\Omega = n\pi$ for an integer $n \geq 1$, which are evident from the drastic drop in displacement amplitude. The zero frequency bandgap is also observed, which is half the width of the rest of them. On the other hand, no zero-frequency bandgap is observed in the second design and the locally resonant bandgaps occur unsymmetrically around the resonant frequencies according to Equation~(\ref{eq:om_R_d2}), which, once again, are evident from the drastic drop in displacement amplitude.}
\label{fig:FRF1}
\end{figure*}

The predictions of the dispersion relations are further verified using the frequency response of a finite structure of twenty unit cells ($\gamma =1,2$ are used as examples) for both EMM designs.~The numerical computation is carried out using MATLAB, and the frequency response function is computed using the finite element method with two-node linear rod elements \cite{petyt2010introduction}.~\HA{After assuming harmonic motion and in the absence of external forcing, the equation of motion is compactly expressed in frequency domain using the following matrix form:
\begin{equation}
    [\mathbf{K}(\omega) - \omega^2 \mathbf{M}] \mathbf{u} = \mathbf{0}
    \label{eq:EOM_FE}
\end{equation}
where $\mathbf{M}$ and $\mathbf{K}$ are the mass and stiffness matrices from the finite element method, while $\mathbf{u}$ is the displacement vector. The frequency-dependent stiffness matrix $\mathbf{K}(\omega)$ incorporates the frequency-dependent elastic foundation stiffness based on Equations~(\ref{eq:kd1}) and (\ref{eq:kd2}).~The structures are base-excited at one end, and Equation~(\ref{eq:EOM_FE}) is rearranged to account for such form of excitation (See Ref.~\cite{Nouh2014} for more details). After computing the degrees of freedom, the amplitude of the displacement near the other fixed end is measured at a swept range of frequencies.~The result of this process is the frequency response functions shown in Figure~\ref{fig:FRF1}, with normalized displacement and frequency.}~The bandgaps are discerned as regions of drastic decrease in displacement amplitude, depicted as the shaded regions in Figure~\ref{fig:FRF1}. As can be seen from the figure, it is evident that the first EMM design has symmetric bandgaps around the resonant frequencies $\Omega = n\pi$ (for integer $n\geq1$), and that the zero-frequency bandgap is half the width of the rest of them with the occurrence of a zero-frequency anti-resonance. Analogously, the frequency response function for the second EMM design evinces that anti-resonances occur according to Equation~(\ref{eq:om_R_d2}), yielding identical bandgaps, yet generally unsymmetrical about the anti-resonance frequencies. Finally, the lack of zero-frequency bandgap is also evident in the second design.

\section{Concluding remarks}
Two designs of active elastic metamaterials (EMM) are proposed, which exhibit an infinite number of locally resonant bandgaps that are evenly spaced and identical in width (except for zero-frequency bandgaps in the first design). Both designs are achieved via a grounded elastic element with a frequency-dependent stiffness.~With the aid of the transfer matrix method, the function describing the stiffness element is carefully designed based on an engineered dispersion relation with an infinite number of local-resonance frequencies.~The dispersion analyses and bandgap predictions are verified using a finite-element-based frequency response function of a finite array of the active EMMs, showing excellent agreement.~\HA{One limitation of the proposed designs is the challenge of experimentally realizing the frequency-dependent stiffness of the elastic support, especially if a specific frequency function is desired.~Nonetheless, the proposed design methodology could be further utilized to engineer new dispersion behaviors in the future.}

\newpage
\section*{References}

\begin{multicols}{2}
\footnotesize
\printbibliography[heading=none]
\end{multicols}
\end{document}